TITLE

Experimental method to determine specific heat and transition enthalpy at a first-order phase transition: fundamentals and application to a Ni-Mn-In Heusler alloy

F. J. Romero (fjromero@us.es), M.C. Gallardo (mcgallar@us.es), J.M. Martín-Olalla (olalla@us.es), J. del Cerro (delcerro@us.es).

Departamento de Física de la Materia Condensada. Facultad de Física. Universidad de Sevilla. Avenida Reina Mercedes s/n 41012 Sevilla SPAIN

Corresponding author: F.J. Romero (fjromero@us.es)

ABSTRACT

A new method that characterizes thermal properties during a first-order phase transition is described. The technique consists in exciting the sample by a series of constant frequency thermal pulses which one in every N pulses --N is a small number like four—being exceedingly large in amplitude. This pulse induces phase transformation which is inhibited during the following smaller pulses due to thermal hysteresis. That way the specific heat for a given mixture of phases can be determined. The results obtained are independent of experimental parameters like the rate and the amplitude of the pulses, unlike what happens in other calorimetric techniques. The method also provides the enthalpy excess by analysing the energy balance between the dissipated heat and the heat flowing during each pulse of measurement.

The protocol is tested to analyse the phase transitions of a Heusler alloy $Ni_{50.54}Mn_{33.65}In_{15.82}$. The paramagnetic-ferromagnetic transition for the austenite phase is continuous and the specific heat shows a lambda anomaly. The martensitic phase transition shows a first-order character and the specific heat follows a step-like behaviour.

1. INTRODUCTION

The measure of thermal properties such as specific heat, enthalpy and entropy is paramount in many topics of physics, chemistry or materials science[1] like the characterization of phase transitions. Thermal properties integrate every contribution to the free energy of the system and thus provide a unique information of the system.

First-order phase transitions are characterized by a few defining properties. They ideally happen at a given temperature. Energy exchange does not cause the sample to heat or cool (sensible heat), instead it causes phase transformation (latent heat) altering the molar fraction of coexisting phases[2]. Thermal properties like entropy and enthalpy excesses show a discontinuity while the specific heat has a singularity at the transition point since the exchange of energy does not cause a change of temperature. An additional feature is the presence of thermal hysteresis between cooling and heating runs.

In real systems this sharp behaviour is usually smeared: temperature gradients, internal stresses, sample inhomogeneities, defects and the like make the transition to happen over an extended range of temperature. The latent heat spreads over that temperature interval and there is also a variation of the specific heat since some fraction of the sample is modifying its temperature.

As a result of all this, the measure of thermal properties during a first-order phase transition is always challenging.

Specific heat c is a primordial property for the thermal characterization of a sample. By integrating the specific heat over temperature, changes of enthalpy over a range of temperature are evaluated. Likewise by integrating c/T, changes of entropy are determined. In the simplest case specific heat is measured by exciting the sample with a thermal pulse; the ratio of the pulse to the change of temperature evaluates the specific heat as per its raw definition.

In first-order phase transitions the experimentalist must discriminate which portion of the heat pulse caused phase transformation (latent heat) and which portion, if any at all, caused a change of temperature (sensible heat). Both contribute to the total change of enthalpy over an extended range of temperature.

Specific heat data obtained in presence of latent heat may not be reproducible because they could depend on the experimental setup including parameters like the amplitude of the heat pulses and the temperature variation rate.

Adiabatic calorimetry is an excellent tool for measuring heat capacity data. However, an intrinsic inaccuracy in the values obtained in the vicinity of a first-order phase transition has been discussed [3]. During a first order phase transition the sample ceases to be a passive element whose temperature increases as energy is dissipated by the heater. Instead, the sample is able to absorb/release energy without increasing/decreasing its temperature. Therefore, the energy balance by which specific heat is determined in adiabatic calorimetry is compromised.

Although high ac calorimetry [4] has a very good temperature resolution, it has been extensively used to study second-order phase transition because it measures heat capacity and it is unable to analyse total enthalpy [5]. On the contrary, differential scanning calorimetry and differential thermal analysis [1] are able to compute total enthalpy changes although the temperature resolution is not good as a consequence of the fast scan rates. However, it is hard to discriminate specific heat contribution and latent heat contributions. Temperature modulated differential scanning calorimetry (TMDSC) combines ac calorimetry and DSC and the temperature ramp is modulated by an alternative component [6,7] to separate the reversing and the non-reversing contributions to the specific heat.

It must be also noticed that the application of commercial heat-flow sensors (Peltier cells) has allowed to build new calorimeters especially suitable to determine the heat capacity and the magnetocaloric effect in first-order phase transitions. The design allows to miniaturize the assembly by using large figure of merit semiconductor thermopiles [8,9]. Microcalorimetry has also been used to study small distributed latent heats and to separate it from the specific heat by using membrane-based microcalorimeters [10].

Relaxation calorimetry has been shown to be an excellent technique to determine heat capacities [11]. The growing availability of commercial, fully automated relaxation calorimetry (specifically the physical property measurement system PPMS from Quantum Design QD) has allowed to extend this type of calorimetry. The standard method, consisting in fitting the heating and cooling branches obtained when the temperature of the sample is modified by a heat pulse, is not appropriate for first-order phase transitions [12]. Different approaches have been done to study first-order phase transitions. The single pulse method consists in applying a temperature pulse large enough to wholly cross the phase transition [13]. For materials with transitions extended over dozens of degrees this method cannot be applied. A different method (successive heating pulses) has been described [14] where a series of short heating pulses (temperature increment lower than 2K) is applied at equally spaced temperatures (typically 2K) across the transition interval, preceded by a long-time stabilization before the power is applied. The key point is that there should not be overlap between the temperatures reached in two consecutive pulses.

We previously developed a method called Squared Modulated Differential Thermal Analysis (SMDTA) which is able to characterize continuous (second-order) and discontinuous (first-order) phase transitions[15,16] even in the neighborhood of a tricritical point where the challenges grow because the specific heat is expected to diverge. In this technique the specific heat is measured by exciting the sample with a series of identical square thermal pulses superposed to a linear temperature ramp.

To allow a more precise study of first-order phase transitions a procedure was developed which consists in combining the results of two experiments on the same sample and under similar temperature variation rates. In the first experiment, the specific heat was determined with aforementioned method. Then a second run recorded the fluxmeter signal under the same linear temperature ramp with no other external excitation, which is a high sensitive differential thermal analysis.

Later, we showed that the analysis of the data from the transient response of the fluxmeters during the specific heat measurement (with the sample being excited by the series of thermal pulses) allowed to obtain a signal which is equivalent to the DTA trace of the second experiment[16]. Therefore, the second experiment became unnecessary. However, a challenge remained: discriminating the specific heat and latent heat during a first order phase transition due to the influence of the latent heat on the transient response during the measurement.

This paper introduces an improvement of the SMDTA technique that allows the measurement of the heat capacity during a first-order phase transition. Unlike the original method, the new technique employs amplitude-modulated thermal pulses, with one in every N pulses being exceedingly large. We take advantage of the thermal hysteresis that follows the large thermal pulse and which inhibits subsequent phase transformations. The fundamental of the method could be applied to other types of calorimetry. For testing the method we choose a Heusler type alloy ($Ni_{50.54}Mn_{33.65}In_{15.82}$) which exhibits a second-order para-ferromagnetic phase transition followed by a first-order martensitic phase transition[17].

2. EXPERIMENTAL DETAILS

The experimental device has been described elsewhere[18]. The sample is sandwiched between two heat fluxmeters, rigid enough to support the sample and to apply, if necessary, a controlled uniaxial stress on the sample. The other end of each fluxmeter is in thermal contact with a calorimetric block. The thermal stability of the equipment allows modifying the temperature of the calorimetric block at rates of 0.02 K/h-0.30 K/h. Two heaters are placed between the sample and the fluxmeters, which can be used to supply heat pulses, whose effect is superposed to the temperature ramp. Heat pulses are able to achieve a temperature amplitude ranging from 0.01K to 0.25K.

The sample was nearly octagonal with flat faces of area 90 $mm^2$ and 1.89mm height. The mass of the sample was 1.5487 g. Sample composition was determined as $Ni_{50.54}Mn_{33.65}In_{15.82}$ from energy dispersive X-ray fluorescence measurements (spectrometer EDAX μFRXEAGLE III) at the Research, Technology and Innovation Centre (CITIUS, University of Sevilla, Spain). The electrons per atom $e/a$ was calculated as the concentration-weighted average of the valence electrons, which are the electrons in the external s, p and partially occupied d orbitals of the constituents. The value of $e/a$ for this sample is 7.883.

3. METHOD

The specific heat of the sample is measured using the following procedure [19]. At time t=$t_0$ the heaters begin to supply a constant power $W_0$ for a time $\tau$ long enough to let the sample reach a steady state temperature distribution, characterized by a temperature difference $\Delta T_1$ between the sample and the calorimetric block. At time $t_1=t_0+\tau$, the power $W_0$ is cut off and the sample relaxes to a new steady state temperature distribution at time $t_2=t_0+2\tau$. The electromotive force given by the fluxmeters *V(t)* is measured during the described process. V(t) is proportional to proportional to the heat flux ϕ(t) crossing the fluxmeters, $\phi(t) = V(t)/\alpha$, where $\alpha$ is the sensitivity of the fluxmeters, determined by calibration.

Figure 1 shows an ideal square pulse (top) and the transient response *V(t)* by the fluxmeters scaled by the highest deviation $V_1$-$V_0$. Three normalized areas can be defined:

$$A_w = \int_{t_0}^{t_0+\tau} \frac{V-V_0}{V_1-V_0} dt \tag{1}$$

$$A_d = \tau - \int_{t_0}^{t_0+\tau} \frac{V-V_0}{V_1-V_0} dt \tag{2}$$

$$A_r = \int_{t_0+\tau}^{t_0+2\tau} \frac{V-V_0}{V_1-V_0} dt \tag{3}$$

The integral of the heat flux over the duration of one cycle amounts to the total energy supplied by the heaters Q=$W_0$ $\tau$.

$$\int_{t_0}^{t_0+2\tau}(\phi - \phi_0)dt = \int_{t_0}^{t_0+2\tau} \frac{V-V_0}{\alpha} dt = \frac{(V_1-V_0)(A_w+A_r)}{\alpha} = W_0\tau = Q \tag{4}$$

Two values of the specific heat are obtained, $c_d$ while the heaters are dissipating (dissipation branch) and $c_r$ while there is no dissipation (relaxation branch) [15]:

$$mc_d = \frac{2}{\beta}(A_d - A_d^0) \tag{5}$$

$$mc_r = \frac{2}{\beta}(A_r - A_r^0) \tag{6}$$

where *m* is the mass of the sample, $A_d^0$ and $A_r^0$ are the areas when no sample is attached to the fluxmeters and $\beta$ is the thermal resistance of the fluxmeters.

When no kinetic process is present, dissipation and relaxation branches are equivalent processes since the temperature of the sample changes equally in both branches and, consequently, the values of $A_d$ and $A_r$ coincide and, hence, $c_d$=$c_r$.

During a first order phase transition the relaxation branch and the dissipation branch may not be equivalent to each other as the latent heat is being absorbed or released. In this case, $c_d$ and $c_r$ could not coincide and they could even show non-regular behavior due to the influence of the latent heat. This fact has been used to ascertain the temperature interval where a sharp first-order phase transition takes place [15].

Due to the large thermal inertia of the calorimetric block this protocol seldom operates in this equilibrium configuration. Instead it operates while the temperature of the calorimetric block is steadily varying at a low rate. Figure 2 shows (panel a and b) a series of thermal pulses and responses; panel (c) shows the evolution of sample temperature with time when the temperature of the calorimetric block is steadily increasing.

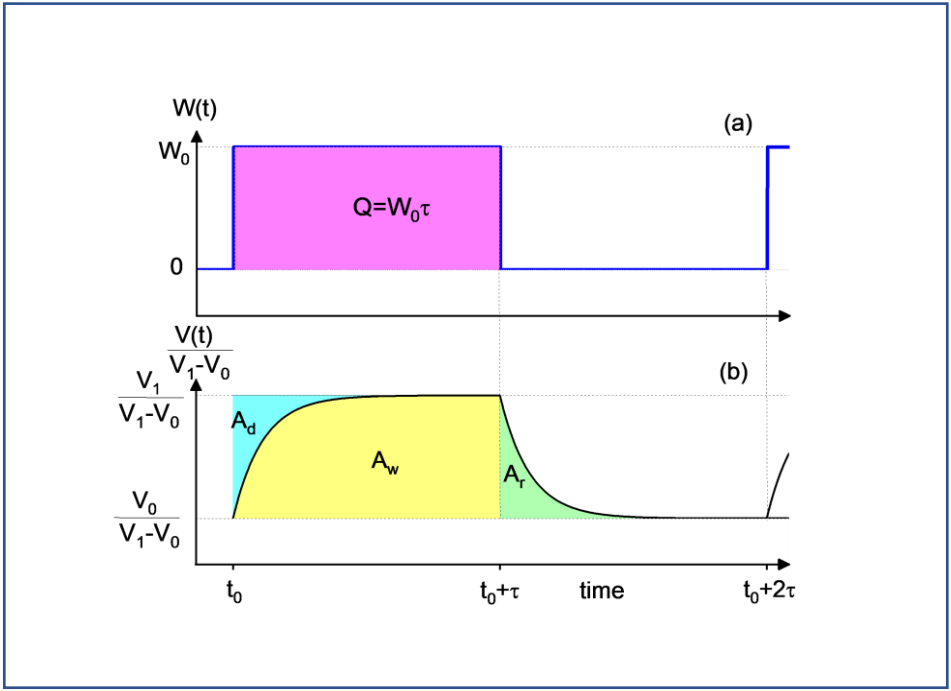

Figure 1: (a) A square thermal pulse of amplitude $W_0$ and duration $\tau$ dissipating $Q=W_0\tau$ and (b) the response given by the fluxmeters scaled by the highest deviation $V_1$-$V_0$. The areas $A_d$, $A_w$ and $A_r$ (see Equations (1), (2) and (3)) are shaded.

When the temperature of the sample reaches a first-order phase transition the thermal pulse will trigger the phase transformation into the high temperature phase. Now energy is not stored as an increase of temperature (sensible heat) but as a structural transformation (latent heat). This will modify the response of the fluxmeters[16,20] and Eqs. 5 and 6 from which specific heat is extracted do not hold.

Also in this case, the difference between the integral of the heat flux and $Q$ is the enthalpy change exchanged in this cycle $\Delta H_i$.

$$\Delta H_i = W_0\tau - \int_{t_0}^{t_0+2\tau}(\phi - \phi_0)dt = W_0\tau - \frac{1}{\alpha}\int_{t_0}^{t_0+2\tau}(V - V_0)dt \qquad (7)$$

Notice that in any of the cycles where the temperature of the sample is within the coexistence interval, the sample will be partially transformed to the high temperature phase since the temperature steadily increases with every cycle ($T_{i+1}>T_i$, see Figure 2c). This fact will happen irrespective of the temperature variation range or the applied heat power.

The new method we present in this work is based in a train of N pulses of which the first one (A) is exceedingly large than the remaining pulses (B), see Figure 2d. During a first-order phase transition pulse A will transform a fraction of the sample into the new high temperature phase but during the relaxation process thermal hysteresis will inhibit the backward transformation.

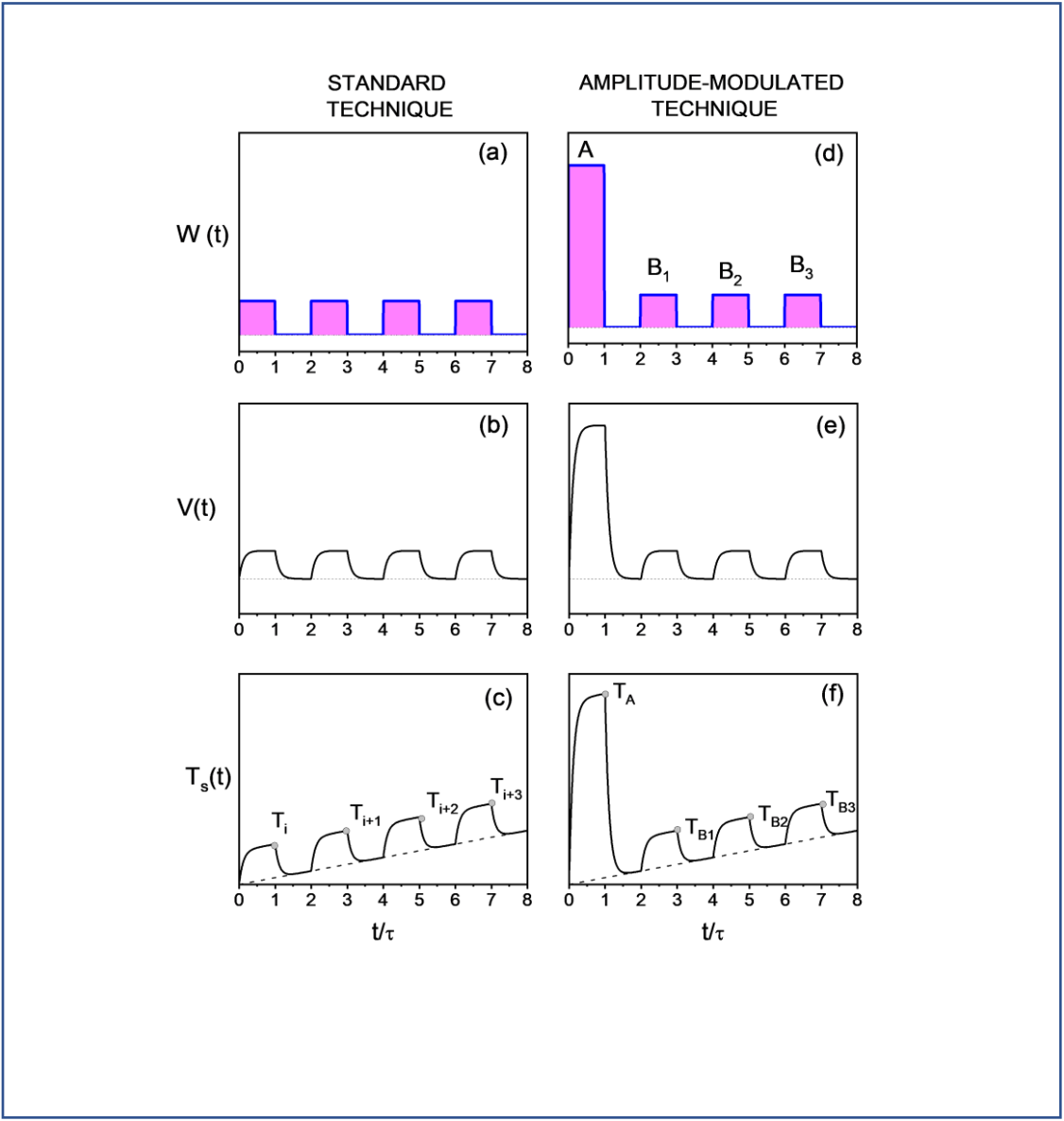

Figure 2. On the left a schematic of the standard technique is represented: a train of square heat pulses with equal amplitude and period (a); the associated response given by the fluxmeters (b); and change of temperature in the sample if a stationary rate of temperature change is set on the calorimetric block (c). Note that $T_{i+1}>T_i$ always. On the right the schematic of the new protocol is represented: a train of square heat pulses of equal period but with modulated amplitude, the first out of four pulses is exceedingly large amplitude (d); the associated fluxmeter signal (e) and the expected change of temperature in the sample (f). Note that the parameters of the experiments are set so that $T_{Bi}<T_A$.

Thereafter pulse B will not ideally induce new phase transformation since the temperature of sample does not exceed the maximum temperature reached during pulse A ($T_{Bi}<T_A$, see Figure 2f). The transformation resumes only when the following pulse A is onset.

In this scenario the enthalpy excess can be determined from the energetic balance of the cycles. On the other hand reproducible specific heat values can be obtained from Eq. 5 and 6 only for B-pulses. The normalized areas of pulses A are not related to specific heat since phase transformation occurs in these cases and fluxmeter responses are distorted [16,20] as energy is structurally stored.

These hypotheses will only be sound as long as the amplitude of pulse A, the amplitude of pulse B and the rate of change of temperature are appropriately set in the way that the temperature of the sample does not exceed maximum temperature reached on pulse A ($T_{Bi}<T_A$, see figure 2f). We recall that the phase transition takes place over an extended range of temperature due to internal stresses, inhomogeneities or defects, to name a few.

For each pulse, Equation (7) is computed. For the first pulse it is expected that $\Delta H_i \neq 0$ if the transition is first-order. If the parameters are set appropriately for the following N-1 pulses or when no phase transition is occurring $Q_i = \tau W_i$ and the integral should match to each other within the experimental resolution.

The transition enthalpy will be calculated from $\Delta H_i = \sum_i \Delta H_i$ and the coexistence interval will be obtained by analysing the temperatures at which the condition $\Delta H_i \neq 0$ has been fulfilled.

4. RESULTS AND DISCUSSION

To test the method a sample of the metamagnetic shape memory alloy Ni-Mn-In was chosen. According to the phase diagram of $Ni_{50}Mn_{50-x}In_x$ alloys [21], a sample of composition $Ni_{50.54}Mn_{33.65}In_{15.82}$ undergoes two phase transitions which are close to each other and near room temperature. On heating there is a first-order phase transition between a paramagnetic-martensitic phase and a ferromagnetic-austenitic phase followed by a second-order phase transition between ferromagnetic-austenitic and paramagnetic-austenitic and phases. The experimental procedure can be applied on both transitions and in the same run without altering the experimental conditions.

In a first run, we recorded the specific heat using the standard procedure[16,19] which consists of a series of identical square thermal pulses. We measured the time constant (78 s) and selected $\tau$=720 s. The first run was performed under a temperature scanning rate r=0.18 K/h ($5.5 \times 10^{-5}$ K/s) and the power dissipated (*W*=0.58mW, *Q*=0.42J) in the heaters was selected to produce a temperature increment $\Delta T$=0.035 K at the end of the dissipation branch. A lambda-like anomaly was observed around 310 K associated to the ferro-paramagnetic (FP) transition of the austenite phase. Despite the compositional variation, the signature is similar to the anomaly seen for other members of the same family [22,23]. At lower temperatures, a strong peak was observed around 288 K which is associated to the austenitic-martensitic (M) phase transition. Sharp peaks have been also observed for the martensitic transition of Ni-Mn-In, Ni-Mn-In-Co or Ni-Mn-In-Sn [17,22–26].

Except around the M transition $c_d$ and $c_r$ match to each other within experimental resolution. We have shown previously that a difference between $c_d$ and $c_r$ is an evidence of a first-order phase transition[15]. Martensitic phase transitions are known to show athermal behaviour, so that there is a distribution of transition temperatures and the latent heat spreads over a wide temperature interval. Scattering of $c_d$ and $c_r$ is likely to be due to the intermittent dynamic of the martensitic transformation which produces quasi-instantaneous heat flux peaks of different amplitudes and durations [27–31].

To shrink the fraction of sample that is transformed during each cycle we reduced the dissipated power (*W*=0.15mW, *Q*=0.11J, $\Delta T$ =0.01K) as well as the temperature variation rate $r$=0.04 K/h (1.1 x$10^{-5}$ K/s) in a second run. The results are also shown in Figure 3. Data for the first and second run coincide over all the temperature interval except for the martensitic phase transition. There is no influence of the measuring parameters on the second order phase transition around 310K. However, during the M transformation the second run produces an anomaly which is markedly lower and narrower than that observed in the first run. Interestingly $c_d$ and $c_r$ keep showing a noisy distribution, showing that some latent heat is disturbing the specific heat measurement.

Therefore, experimental conditions impact the results. We stress the fact that this is happening even when the temperature scan rate is as low as 0.04 K/h (1.1 x$10^{-5}$ K/s) and the amplitude of the thermal pulse on the sample is as low as 0.01 K. The magnitude and the width of the peak depends on the experimental conditions that modify the kinetics of the transformation, even if the experimental device is not altered. Under different kinetics, the molar fraction of each phase evolves in a different way and the results for specific heat data change. Therefore they cannot represent equilibrium values.

In a dynamic experiment, with temperature smoothly varying in time, latent heat and sensible heat are mixed up during a first-order phase transition which causes the measurement method to yield an effective, or generalized, specific heat that blends both contributions. It provides valuable information about the phase transition and would allow the comparison of different samples if the experimental conditions do not change. Nevertheless, a distribution of peak heights can be traced for the martensite-austenite phase transition in Heusler alloys [22, 24-26]. In our set-up it is the right-hand side of Eq (5) and (6) that brings these effective values. Figure 3 shows that our effective specific heat actually depends on the experimental conditions and do not correspond to equilibrium values.

The new method was then tested. A train of N=4 pulses was set with the first pulse (A) producing a temperature increment in the sample of 0.22 K (W=3.85mW, Q=2.80J) and the remaining three pulses (B) being 25 times smaller (W=0.15mW, Q=0.11J, $\Delta$T=0.01K). The scanning temperature rate for this experiment was 0.04 K/h.

Figure 4 shows the results under these conditions. Only around the M transition results for the new method markedly differ from results for the standard method, as we had predicted.

For A pulses the sample is markedly heated during the dissipation branch which, after some temperature is reached, makes the transformation begins. The associated enthalpy modifies the transient response by the fluxmeter and prevents from reaching a stationary state at the prescribed time $\tau$ (the intermittent behaviour of the transformation characteristically induces heat peaks during the dissipation branch. The thermal signature of these jerks extends to the relaxation branch). Dissipation and relaxation specific heats $c_d$ and $c_r$ show a peak which is larger in amplitude and width for $c_d$.

Thermal hysteresis is expected to prevent the reversal transformation after the sample relaxes from pulse A. Notice that Ni-Mn-In compounds typically show hysteresis larger than 1K in the M transformation. Ideally during a B pulse the dissipation branch should not be able to bring new transformations in the sample. Our experimental results fit to this hypothesis: the peak disappears for pulse B data while $c_d$ and $c_r$ match to each other within our experimental resolution.

Figure 4b shows $\Delta H_i$ (see Equation 7) versus temperature. For the FP phase transition, $\Delta H_i$ is zero for all the pulses, irrespective of their amplitude. This is the signature of a second order phase transition.

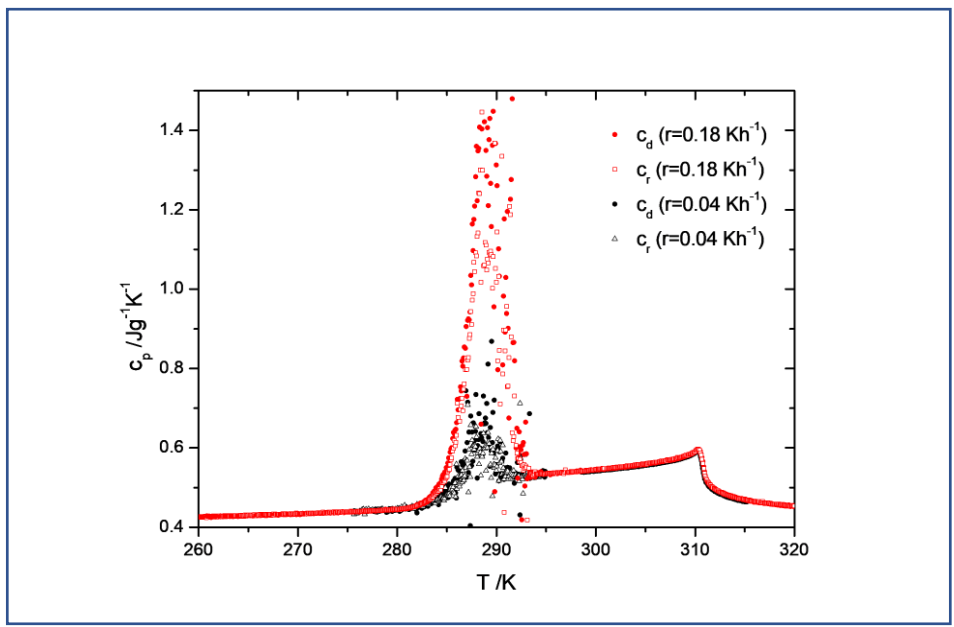

Figure 3: Specific heat data vs temperature in a $Ni_{50.54}Mn_{33.65}In_{15.82}$ sample. Specific heat data for the dissipation branch ($c_d$) and for the relaxation branch ($c_r$) were obtained from Eq. 5 and 6 after a train of identical heat pulses excited the sample. Data in read stands for $r$=0.18 K/h and Q=0.44J; data in black for r=0.04 K/h and Q=0.11J. The reduction in r and Q decreases the peak height but increases the noise.

For the M transition, $\Delta H_i$ is markedly different from zero for A pulses. For B pulses, $\Delta H_i$ is about 0.4% of $\Delta H_i$ for A pulses (see the inset in panel (b)). The magnitude of the anomaly decreases for pulses $B_2$ and $B_3$. It indicates that although there is still a remnant phase transition taking place, the involved fraction of sample is much smaller.

Although the sensitivity of the system allows to measure such small enthalpy changes during B pulses, specific heat data obtained for pulse B are less affected by the latent heat. While for A pulses $m\Delta H_i$ is 9% of Q at most, for B pulse it only amounts to 0.4% of Q at most, which is unable to exclude a null $c_d$-$c_r$ within our experimental resolution.

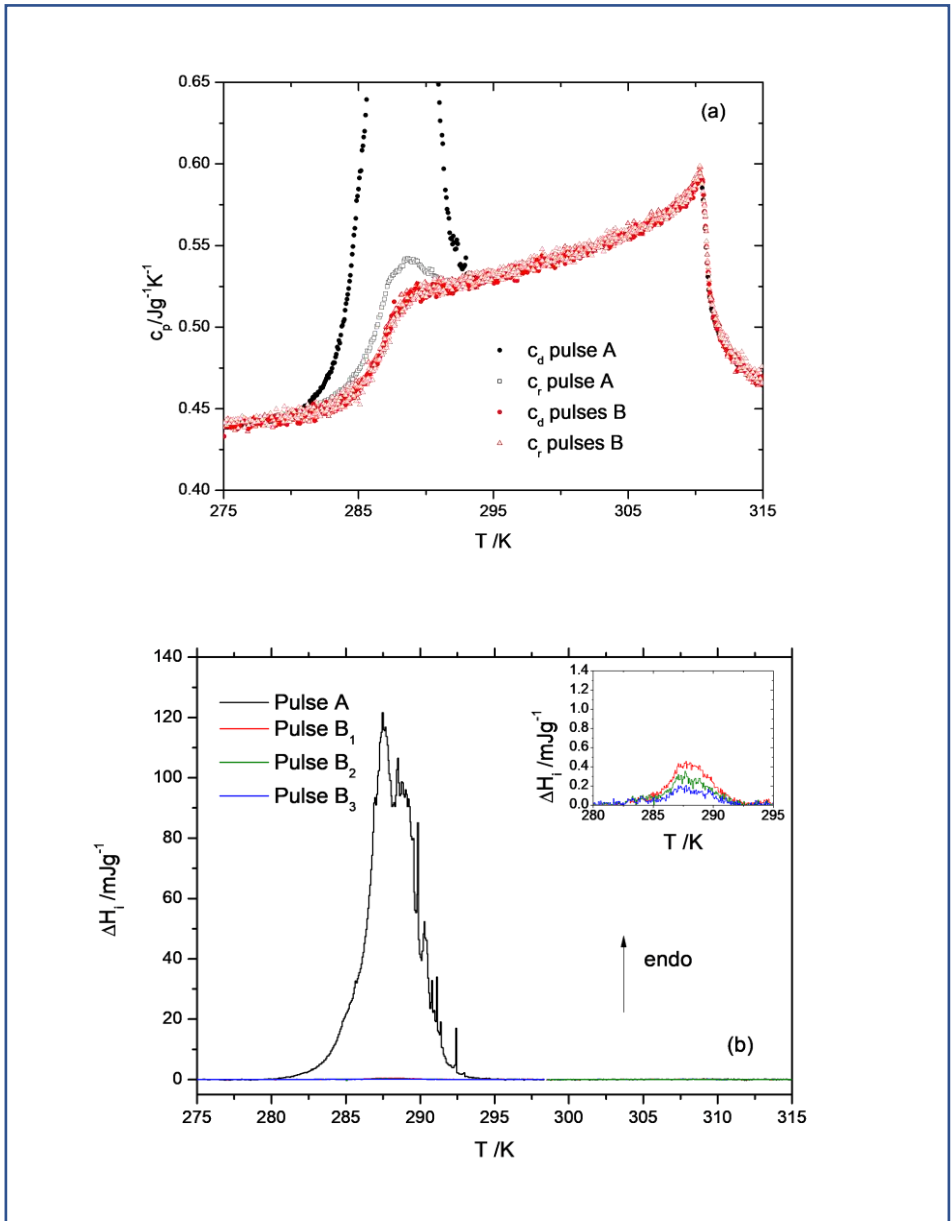

Figure 4: (a) Specific heat data vs temperature in a $Ni_{50.54}Mn_{33.65}In_{15.82}$ sample. Specific heat data were obtained from a train of amplitude modulated heat pulses through Eq. 5 and 6, (b) the transition enthalpy vs temperature in the same run obtained from Eq. 7. The inset in panel (b) enlarges the coexistence interval for pulses B. The vertical axis in the inset spans for 1/100 of the span in the main plot.

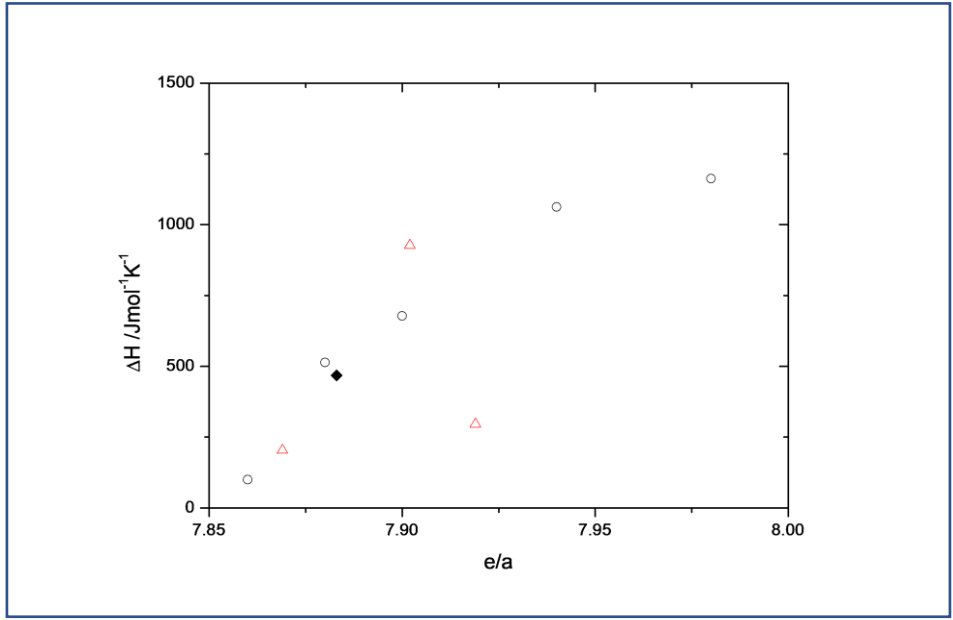

Figure 5: The excess enthalpy for the FP phase transition in different Ni-Mn-In alloys versus e/a: triangles [32], circles [33], diamond (this work).

The anomaly for $c_p$ then looks like step-like transition in agreement with results obtained in a sample of the same family [22]. In this reference a peak was superimposed to the step anomaly. The peak is missed in our data after our technique dramatically reduces the impact of the latent heat in the measurement. The difference between the specific heat for both phases could be related to the different magnetic character of both phases. Measurements of the magnetization would be necessary to ascertain the origin of the difference but this study is beyond the scope of this paper.

The enthalpy jump associated to the M transition is obtained by adding up $\Delta H_i$ for each cycle from 280 K to 295 K, where $\Delta H_i$ is non-zero. This yields $\Delta H$=7.06 J/g or 468 J/g. This result combined with the electronic concentration e/A of the sample is in line with previously reported values for $Ni_{0.50}Mn_{0.50-x}In_x$.[32,33] (see Figure 5).

5. CONCLUSIONS

A new technique able to improve the discrimination of transition enthalpy (energy stored structurally) from specific heat (energy stored kinetically) in first-order structural phase transitions has been designed and successfully tested on a sample of $Ni_{50.54}Mn_{33.65}In_{15.82}$.

The technique uses amplitude-modulated heat pulses during heating scans. The amplitude of one in every four pulses is exceedingly large ($\Delta T \sim 0.2K$) so as to induce phase transformation which stores energy structurally and gives rise to transition enthalpy. In the remaining pulses the amplitude is decreased by a factor 25 so that phase transformation is almost inhibited due to hysteresis and the specific heat for the given molar fraction of phases can be determined.

For the $Ni_{50.54}Mn_{33.65}In_{15.82}$ phase transitions, results show that the ferromagnetic-paramagnetic transition is continuous while the martensitic phase transition is discontinuous. In the latter, the standard procedure gives an effective heat capacity which includes the contributions from the specific heat and the latent heat. The effective heat capacity gives valuable calorimetric information but depends on the temperature variation rate and the heat dissipated during the measurement and cannot be easily related to equilibrium values.

The new protocol represents a progress in the SMDTA technique when it is applied to the study of first-order phase transition with a large thermal hysteresis like in shape memory alloys like the sample studied in this work. The extended thermal hysteresis allows the inhibition of the phase transformation during the smaller pulses, resulting in a better characterization of the transition, since the measured specific heat data are closer to the equilibrium values. This specific heat smoothly changes during the coexistence interval from the austenite value to the martensite value, in contrast with the peak which would is observed under the idea of an effective specific heat that mixes both contributions. The latent heat is also determined from the energetic balance of each pulse.

Should the thermal hysteresis of a phase transition be smaller, a reduction in the temperature amplitude of the pulses and in the rate of temperature change would be mandatory to accomplish the condition $T_a > T_b$. In the case of a first-order phase transition with little thermal hysteresis, the standard protocol could be used.

The large heat capacity of the addenda in our device (which is attributed mainly to the fluxmeters) requires samples above 200 mg and sets a constant time of several tens of seconds. We are now working in a new device based in the miniaturization of the assembly and the thermopiles which would allow to study smaller samples and to reduce the time necessary to perform the measurements.

The amplitude modulated design of heat pulses could be also applied to other calorimetric techniques which take advantage of heat pulses to measure specific heat.

A bold challenge is to adapt the amplitude modulation to cooling scans. This would require to replace the heaters by a Peltier or a thermomagnetic device because in cooling scans phase transformation requires further active cooling.

5. ACKNOWLEDGMENTS

The authors wish to thank Prof. A. Planes (Universidad de Barcelona) for supplying the sample and Dr. R. Cano (Universidad de Sevilla) for his help in cutting the sample.